\documentclass[aps,tightenlines,notitlepage,onecolumn,a4paper,preprintnumbers,superscriptaddress,nofootinbib,11pt]{revtex4-1}
\pdfoutput=1

\usepackage[mathscr]{eucal}
\usepackage{color,amsmath,amssymb,amsfonts}
\usepackage{enumerate}
\usepackage{hyperref}
\usepackage{graphicx}
\usepackage{psfrag}
\usepackage{dcolumn,bm}
\usepackage{mdframed}

\usepackage[utf8]{inputenc}

\newcommand{\bea}{\begin{eqnarray}}
\newcommand{\beal}[1]{\begin{eqnarray}\label{#1}}
\newcommand{\eea}{\end{eqnarray}}
\newcommand{\be}{\begin{equation}}
\newcommand{\bel}[1]{\begin{equation}\label{#1}}
\newcommand{\ee}{\end{equation}}

\newcommand{\bit}{\begin{itemize}}
\newcommand{\eit}{\end{itemize}}
\newcommand{\ben}{\begin{enumerate}}
\newcommand{\een}{\end{enumerate}}

\def\del{\partial}

\newcommand{\eqn}[1]{Eq.~(\ref{#1})}

\newcommand{\eqq}[1]{(\ref{#1})}
\newcommand{\fig}[1]{Fig.~\ref{#1}}

\newcommand{\bal}{\begin{align}}
\newcommand{\eal}{\end{align}}
\newcommand{\bse}{\begin{subequations}}
\newcommand{\ese}{\end{subequations}}

\def\R{{\mathcal{R}}}

\def\E{{\mathcal{E}}}
\def\Peq{{{P}}_{\rm eq}}
\def\Pt{{{P}}_T}
\def\Pl{{{P}}_L}

\begin{document}

\preprint{ICCUB-17-007}

\title{Phase Transitions, Inhomogeneous Horizons and 
Second-Order Hydrodynamics}

\author{Maximilian Attems}
\affiliation{Departament de F\'\i sica Qu\`antica i Astrof\'\i sica \&  Institut de Ci\`encies del Cosmos (ICC), Universitat de Barcelona, Mart\'{\i}  i Franqu\`es 1, 08028 Barcelona, Spain}
\author{Yago Bea}
\affiliation{Departament de F\'\i sica Qu\`antica i Astrof\'\i sica \&  Institut de Ci\`encies del Cosmos (ICC), Universitat de Barcelona, Mart\'{\i}  i Franqu\`es 1, 08028 Barcelona, Spain}
\author{Jorge Casalderrey-Solana} 
\affiliation{Rudolf Peierls Centre for Theoretical Physics, University of Oxford, 1 Keble Road, Oxford OX1 3NP, United Kingdom}
\author{David Mateos}
\affiliation{Departament de F\'\i sica Qu\`antica i Astrof\'\i sica \&  Institut de Ci\`encies del Cosmos (ICC), Universitat de Barcelona, Mart\'{\i}  i Franqu\`es 1, 08028 Barcelona, Spain}
\affiliation{Instituci\'o Catalana de Recerca i Estudis Avan\c cats (ICREA), 
Llu\'\i s Companys 23, Barcelona, Spain}
\author{Miquel Triana}
\affiliation{Departament de F\'\i sica Qu\`antica i Astrof\'\i sica \&  Institut de Ci\`encies del Cosmos (ICC), Universitat de Barcelona, Mart\'{\i}  i Franqu\`es 1, 08028 Barcelona, Spain}
\author{Miguel Zilh\~ao}
\affiliation{Departament de F\'\i sica Qu\`antica i Astrof\'\i sica \&  Institut de Ci\`encies del Cosmos (ICC), Universitat de Barcelona, Mart\'{\i}  i Franqu\`es 1, 08028 Barcelona, Spain}
\affiliation{CENTRA, Departamento de F\'\i sica, Instituto Superior T\'ecnico, Universidade de Lisboa, Avenida Rovisco Pais 1, 1049 Lisboa, Portugal}


\begin{abstract}
We use holography to study the spinodal instability of a four-dimensional, 
strongly-coupled gauge theory with a first-order thermal phase transition. We place the theory on a cylinder in a set of homogeneous, unstable initial states. The dual gravity configurations are black branes afflicted by a Gregory-Laflamme instability. We numerically evolve Einstein's equations to follow the instability until the system settles down to a stationary, inhomogeneous black brane. The dual gauge theory states have  constant temperature but non-constant energy density.  We show that the time evolution of the instability and the final states are accurately described by second-order hydrodynamics. In the static limit, the latter reduces to a single, second-order, non-linear differential equation from which  the inhomogeneous final states can be derived.

\end{abstract}
\maketitle

\section{Introduction}
Hydrodynamics is one of the most successful theories in physics, capable of describing the dynamics of very different  systems over an enormous range of length scales. Traditionally, it is thought of as an effective theory for the conserved charges of a system, constructed as a derivative expansion around local thermal equilibrium. From this perspective hydrodynamics is only expected to be valid when these gradient corrections are small compared to all the microscopic scales.  However, in recent years it has been discovered that the regime of validity is actually much broader.  

From an experimental viewpoint, hydrodynamics has been extremely successful at describing the post-collision dynamics of the drops of matter produced  in 
ultra-relativistic collisions of large nuclei at RHIC 
\cite{Ackermann:2000tr,Adler:2003kt, Back:2004mh} and  the 
LHC \cite{ATLAS:2012at,Chatrchyan:2012ta,Aamodt:2010pa,Aamodt:2010pa}. 
Multi-particle correlations  in these collisions are well described by hydrodynamics~\cite{Huovinen:2001cy,Teaney:2001av,Hirano:2005xf,Luzum:2008cw,Schenke:2010rr,hiranoLHC,Shen:2014vra,
Bernhard:2016tnd}, provided one assumes that the latter is valid in the presence of large gradients  \cite{Romatschke:2016hle}. Also, the recent finding  of similar correlations in even smaller proton-proton collisions \cite{Khachatryan:2010gv,Aad:2015gqa,Khachatryan:2016txc} provides a strong indication that hydrodynamics is applicable even at a baryonic scale \cite{Bozek:2011if, Schenke:2014zha,Kozlov:2014fqa,Weller:2017tsr}. From a theoretical viewpoint, studies of non-abelian gauge theories have shown that hydrodynamics is valid for systems with large gradients both at strong~\cite{Chesler:2010bi,Heller:2011ju,Casalderrey-Solana:2013aba,Chesler:2016ceu,Attems:2016tby} and weak coupling  
\cite{Kurkela:2015qoa,Keegan:2016cpi}. 

In this paper we use the gauge/string duality to test the validity of hydrodynamics in a theory  with a first-order thermal phase transition. 
We place the theory on a cylinder in a variety of homogeneous, unstable initial states. 
In the gauge theory the instability is a spinodal  instability associated to the presence of a first-order phase transition. On the gravity side the instability is a long-wave length, Gregory-Laflamme-type instability \cite{Gregory:1993vy}. We use the gravity description to follow the evolution of these states until they settle down to a stationary, inhomogeneous final state. We then compare both the time evolution and the final configurations to the corresponding hydrodynamic predictions. Recent related work includes \cite{Janik:2015iry,Gursoy:2016ggq,Dias:2017uyv}.

To the best of our knowledge, from the boundary theory viewpoint our results provide the first example of a first-principle, non-perturbative, complete  dynamical evolution of a spinodal instability in a strongly coupled gauge theory. From the gravity viewpoint, they provide the first example of the full dynamical evolution of a Gregory-Laflamme-type instability in an asymptotically anti-de Sitter space.

\section{Model}
\begin{figure}[tbhp]
\begin{center}
\includegraphics[width=.8\textwidth]{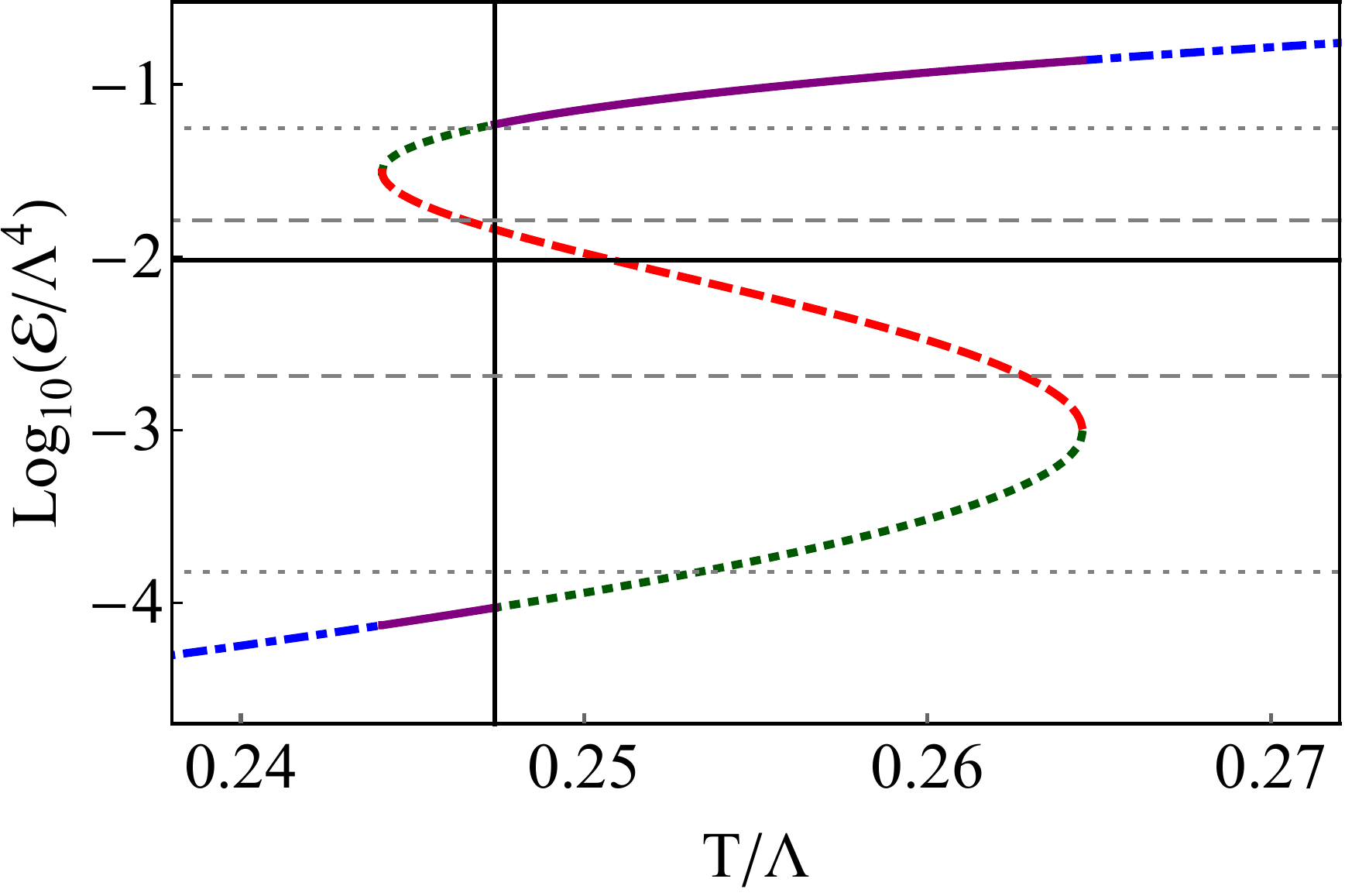}
\caption{\label{fig:energydensity} 
Energy density versus temperature for the gauge theory dual to \eqq{eq:action}. At high and low $T$ there is only one phase shown in dashed-dotted blue. The   preferred phase in the multivalued region  is shown in solid purple. The dotted green curve is metastable. The dashed red curve is locally unstable. The black vertical  line indicates $T_c= 0.247 \Lambda$. The top (bottom) dashed, grey horizontal line indicates the highest (lowest) average energy density that we have considered.  The top (bottom) dotted, grey horizontal line indicates the maximum (minimum) value of the energy in the corresponding final states. The  solid, black horizontal line is the state for which we show specific results.}
\end{center}
\end{figure}

Motivated by simplicity, we study the Einstein-plus-scalar model with action 
\be
\label{eq:action}
S=\frac{2}{\kappa_5^2} \int d^5 x \sqrt{-g} \left[ \frac{1}{4} \R  - \frac{1}{2} \left( \nabla \phi \right) ^2 - V(\phi) \right ] .
\ee
The potential is
\be
\label{eq:pot}
\ell \, V=-3 -\frac{3\phi^2}{2}  - \frac{\phi^4}{3}  - \frac{\phi^6}{3 \phi_M^2} 
+ \frac{\phi^6}{2 \phi_M^4} -\frac{\phi^8}{12 \phi_M^4}  
\ee
with $\ell$ the asymptotic curvature radius and $\phi_M=2.3$. The dual gauge theory is  a Conformal Field Theory (CFT) deformed with a dimension-three scalar operator with source $\Lambda$, which appears as a boundary condition for the scalar. This is exactly the potential of \cite{Attems:2016ugt} except for the sign reversal of the fourth term in $V$. This difference has dramatic implications for the thermodynamics of the gauge theory, which we extract  from the homogeneous black brane solutions of the gravity model. In particular, the gauge theory possess a first-order phase transition at a critical temperature 
$T_c= 0.247 \Lambda$, as illustrated by the multivalued plot of the energy density as a function of the temperature  in \fig{fig:energydensity}. States on the dashed red curve are locally thermodynamically unstable since the specific heat  is negative, $c_v =d \E / dT <0 $. In this region 
 the speed of sound, $c_s^2 = s/c_v$, with $s$ the entropy density, becomes imaginary. This leads to a dynamical, spinodal 
instability (see e.g.~\cite{Chomaz:2003dz}) whereby the amplitude of small sound excitations grows exponentially with a momentum-dependent growth rate dictated by the sound dispersion relation:
\be
\label{eq:growth}
\Gamma (k)\simeq \left | c_s \right |  k - \frac{1}{2 T} \left(\frac{4}{3} \frac{\eta}{s}  + \frac{\zeta}{s} \right)   k^2 \, , 
\ee
where $\eta$ and $\zeta$ are the shear and bulk viscosities. In our model $\eta/s=1/4\pi$ \cite{Kovtun:2004de} and we compute $\zeta$  numerically following \cite{Eling:2011ms}. 

The corresponding statement on the gravity side is that the  black branes dual to the states on the dashed red curve are afflicted by a long-wave length  instability. Although this is similar to the  Gregory-Laflamme  instability of 
Ref.~\cite{Gregory:1993vy}, we will see below that there are also crucial differences between the two.

\section{Inhomogeneous horizon}
\label{horizon}
To investigate  the fate of the spinodal instability we compactify the gauge theory direction $z$ on a circle of length $L\simeq 57/\Lambda$. This infrared cut-off reduces the number of unstable sound modes to a finite number. We then consider a set of homogeneous, unstable initial states with energy densities in the range $\mathcal{E}/\Lambda^4 \simeq (0.002,0.016 )$\footnote{We  work with the rescaled quantities 
$({\cal E}, \Pl, \Pt)=
(\kappa_5^2/ 2 \ell^3) (-T^t_t, T^z_z, T^{x_\perp}_{x_\perp})$.}, as indicated by the grey, dashed  horizontal lines in \fig{fig:energydensity}. For concreteness we will show results for the state with $\mathcal{E}/\Lambda^4 \simeq 0.0096$, whose temperature and entropy density are 
$T_i \simeq 0.251 \Lambda$ and $s_i=0.037\Lambda^3$. To trigger the instability, we introduce a small $z$-dependent perturbation in the energy density corresponding to a specific Fourier mode on the circle. For concreteness we will show results for the case with  $k=3(2\pi /L)\simeq 1.3 T_i$. This mode is unstable with positive growth rate 
 $\Gamma =0.0247 \,\Lambda$ according to \eqq{eq:growth}.
 For numerical simplicity we impose homogeneity along the transverse directions.\footnote{Sound modes with transverse momentum would also be unstable. We thus do not investigate the most general possible end state. We leave this for future work. 
} 

On the gravity side the sound-mode instability may be viewed \cite{Buchel:2005nt,Emparan:2009cs,Emparan:2009at} as a Gregory-Laflamme instability \cite{Gregory:1993vy}.
We follow it by numerically evolving the Einstein-plus-scalar equations as in \cite{Attems:2016tby,Attems:2017zam} until the system settles down to a state with an inhomogeneous Killing horizon with constant temperature $T_f = 0.250 \Lambda$. Note that this is close but not identical to  $T_i$ or $T_c$. 
\begin{figure}[tbhp]
\begin{center}
\includegraphics[width=.8\textwidth]{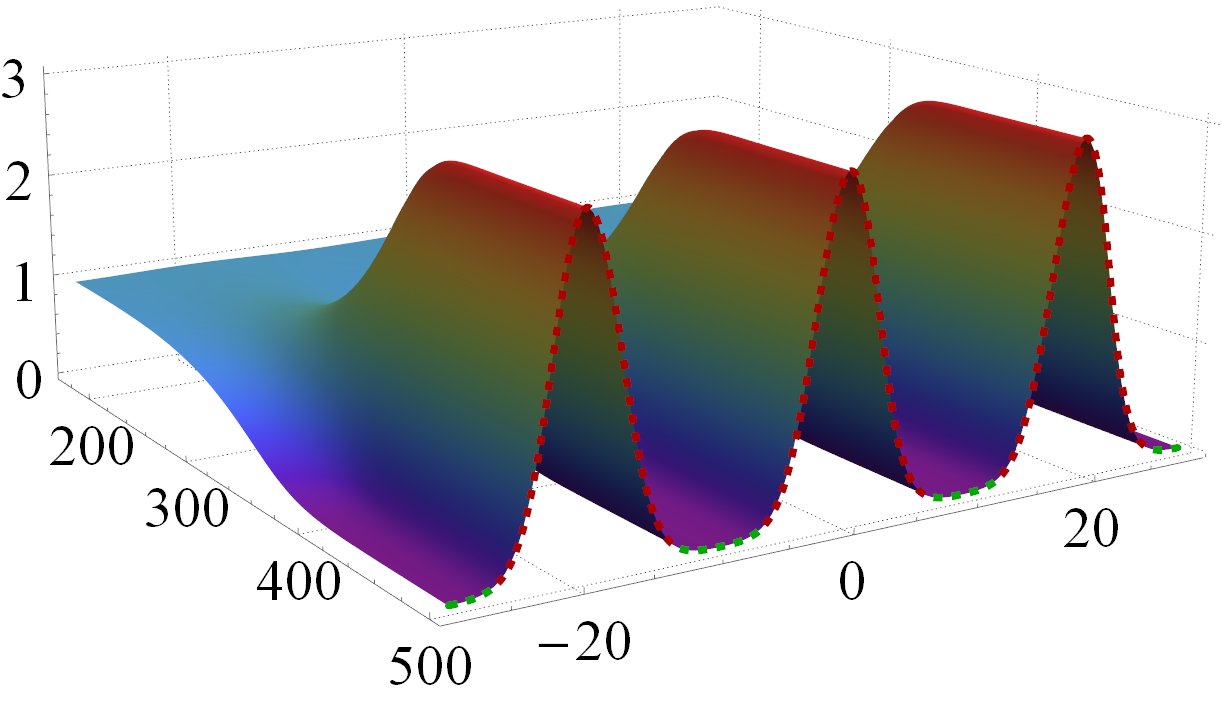}
 \put(-400,210){\Large $ 10^{2} \E / \Lambda^4$} 
  \put(-350,35){\Large $t \Lambda$} 
 \put(-120,10){\Large $z \Lambda$} 
\caption{\label{3Denergy} Energy density for the initial state indicated by the solid, black horizontal line in \fig{fig:energydensity}, perturbed by the third Fourier mode. The color coding on the final-time slice is the same as in \fig{fig:energydensity}.}
\end{center}
\end{figure}
\begin{figure}[tbhp]
\begin{center}
\includegraphics[width=.8\textwidth]{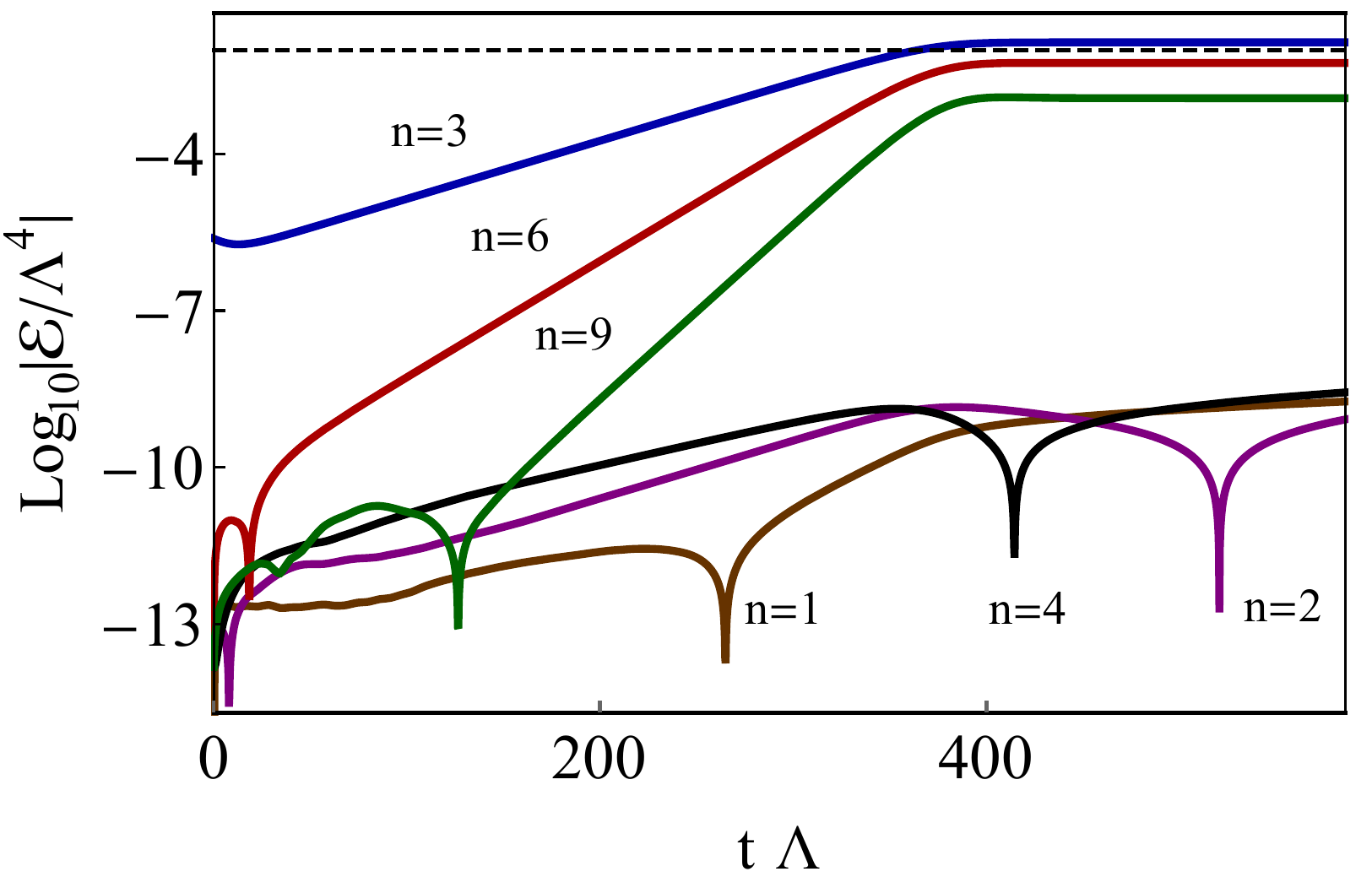}
\caption{\label{fig:modes} Time evolution of some Fourier modes of the energy density $\E$. The dashed horizontal line is the average energy density.
 }
\end{center}
\end{figure}
From the dynamical metric  we extract the boundary stress tensor. The result for the energy density is shown in \fig{3Denergy}. 
The time dependence of the amplitudes of several 
 modes of the energy density is shown in \fig{fig:modes}. The $n=3$ mode grows with a rate that agrees with \eqq{eq:growth} within 4\%. Resonant behavior makes the modes with $n=6, 9, \ldots$  grow at a rate  that is roughly the corresponding multiple of the $n=3$ rate. Numerical noise makes some non-multiples of the $n=3$ mode (of which only three are shown in the figure) grow too.
At late times the non-linear dynamics stops the growth of all these  modes and the system settles down to a stationary, inhomogeneous configuration consisting of three identical domains. The fact that their size is comparable to the inverse temperature,  
$\Delta z = L/3 \simeq 0.4/T_f$, is our first indication that spatial gradients are large.

\begin{figure}[tbhp]
\begin{center}
\includegraphics[width=.7\textwidth]{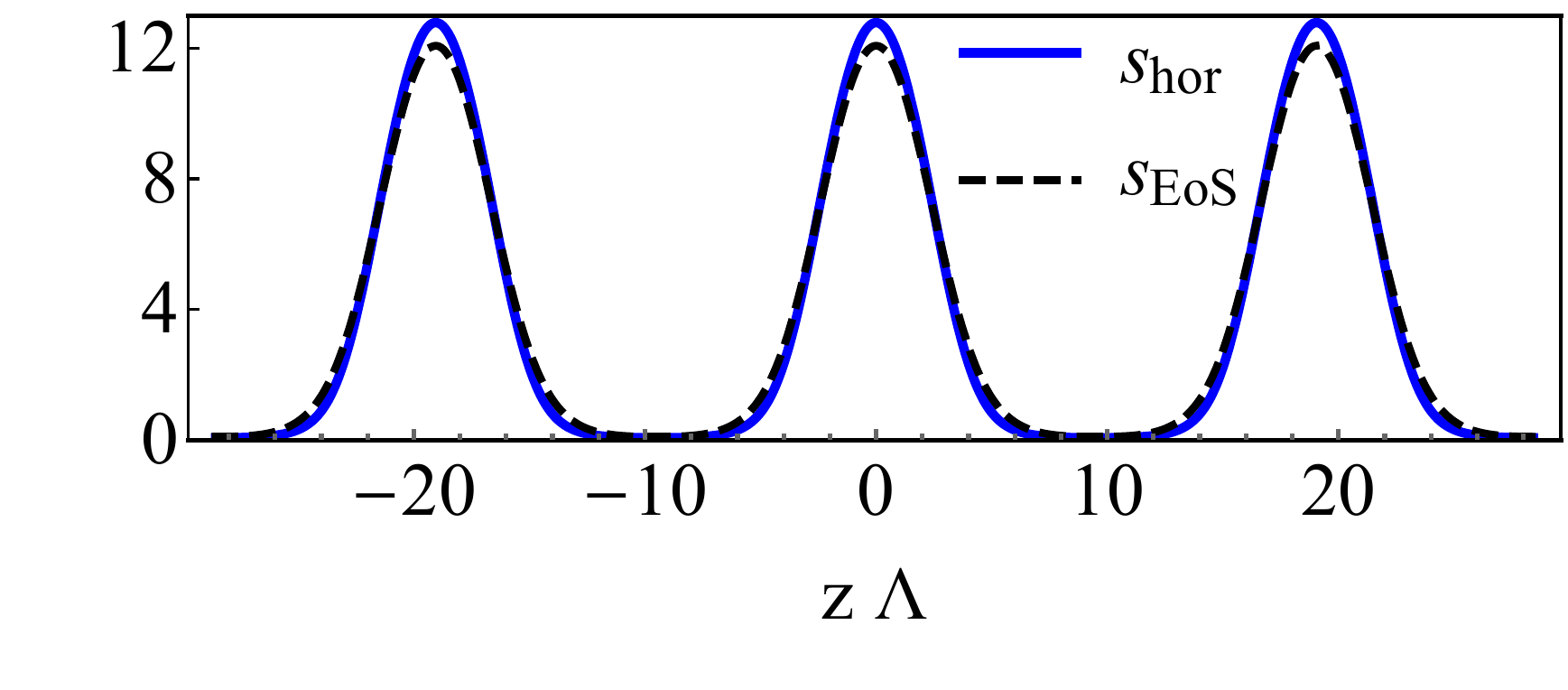}
\caption{\label{fig:bumps} Final entropy density extracted from the area of the horizon (continuous, blue curve) and estimated from the equation of state (dashed, black curve), in units of $\Lambda^3/100$.  
}
\end{center}
\end{figure}

 In \fig{fig:bumps} we show the final entropy density as extracted from the area of the horizon. The fact that $s$ is not constant proves that the horizon itself (not just the boundary energy density) is  inhomogeneous. We see in the figure that $s$ is very well estimated by a point-wise application of the equation of state to the final energy density of \fig{3Denergy}, suggesting that the evolution is quasi-adiabatic. This is confirmed by the fact the final average entropy density is only $1\%$ larger than the initial one. 

In the paragraphs above we have been careful to use the term ``stationary'' as opposed to ``static'' to refer to the final, time-independent state to which the system settles down. The reason is that the final metric is  stationary but not static. The physical reason for this is that the $dt dz$ off-diagonal terms in the final metric do not vanish. As a consequence,  the final state is time-independent but not time-reversal invariant (although it is invariant under the simultaneous sign reversal of both $t$ and $z$). The fact that this is a true property of the solution and not just a consequence of our metric ansatz follows from the fact that the timelike Killing vector of our solution is not hypersurface-orthogonal. Interestingly, the $g_{tz}$ component of the final-state metric falls off quickly enough near the AdS boundary so that there is no energy flux in the $z$-direction. In other words, the one-point function of the $T^t_z$ component of the boundary stress tensor vanishes identically. As a consequence, at the level of one-point functions the final state on the gauge theory side looks not only stationary but static. Since hydrodynamics is an effective theory for one-point functions, in the next section we will speak of a (hydro)static final state. However, it must be kept in mind that the full microscopic state in the gauge theory is only stationary, since $n$-point functions with $n>1$ would be sensitive to the properties of the dual metric deep in the bulk and hence would reveal the lack of staticity.

\section{Hydrostatic final state}
Previous holographic studies have established that hydrodynamics can describe the evolution of the stress tensor in the presence of large gradients.
%
 Here we investigate whether it can also describe the static inhomogeneous configuration at the endpoint of the spinodal instability.

In the hydrodynamic approximation  the stress tensor is expanded in spatial gradients in the local fluid rest frame. Up to second order, the constitutive relations are 
\be
\label{eq:hydroT}
T^{\rm hyd}_{\mu \nu} = T^{\rm ideal}_{\mu \nu} - \eta \, \sigma_{\mu \nu} -
\zeta \, \Pi \, \Delta_{\mu \nu} + \Pi_{\mu \nu}^{(2)}
\ee
where, in the fluid rest frame, $T_{\mu \nu}^{\rm{ideal}}={\rm Diag}\{ \E, \Peq (\E) \}$, $\Peq (\E)$ is the equilibrium pressure of the homogeneous states shown in \fig{fig:energydensity}, $\sigma_{\mu \nu}$ and $\Pi$ are the shear and bulk stresses, and $\Delta_{\mu \nu}$ is the projector onto the spatial directions.   
The tensor $\Pi_{\mu \nu}^{(2)}$ contains all the second-order terms. In a non-conformal theory this tensor contains thirteen $\E$-dependent second-order transport coefficients and its explicit expression may be found in \cite{Romatschke:2009kr}.
A subset of these coefficients for the model of \cite{Attems:2016ugt}  has been computed in \cite{Kleinert:2016nav}.

In a static configuration the  fluid three-velocity field is identically zero, the stress tensor is diagonal, both $\sigma_{\mu \nu}$ and $\Pi$ vanish, 
and the leading gradient corrections are those in $\Pi_{\mu \nu}^{(2)}$. This  tensor  also simplifies since only two independent second-order terms survive in this limit. 
In the Landau frame, the constitutive relations reduce to 
\bea
\label{hydroap1}
\Pl^{\rm hyd}&=& \Peq (\E) + c_{\rm L}(\E) (\del_z \E)^2 +   f_{\rm L}(\E) (\del^2_z \E)\,,
\\
\label{hydroap2}
\Pt^{\rm hyd}&=& \Peq (\E) + c_{\rm T}(\E) (\del_z \E)^2 +   f_{\rm T}(\E) (\del^2_z \E)\,.
\eea 
The statement that hydrodynamics describes the final states is the statement that the corresponding pressures are given by these equations with second-order transport coefficients $c_{\rm L, \,  T} (\E)$ and $f_{\rm L,\, T} (\E)$  that depend on the local energy density\footnote{They can be expressed as linear combinations of the coefficients 
 $\tau_\pi \eta$, $\tau_\Pi \zeta$, $\lambda_4$ and $\xi_4$ identified 
 in \cite{Romatschke:2009kr}.} but \emph{not} on any other details of the final states. We have verified this state independence by varying both the average energy densities within the range shown in \fig{fig:energydensity} and the length of the circle $L$. 

\begin{figure}[tbph]
\begin{center}
\includegraphics[width=.7\textwidth]{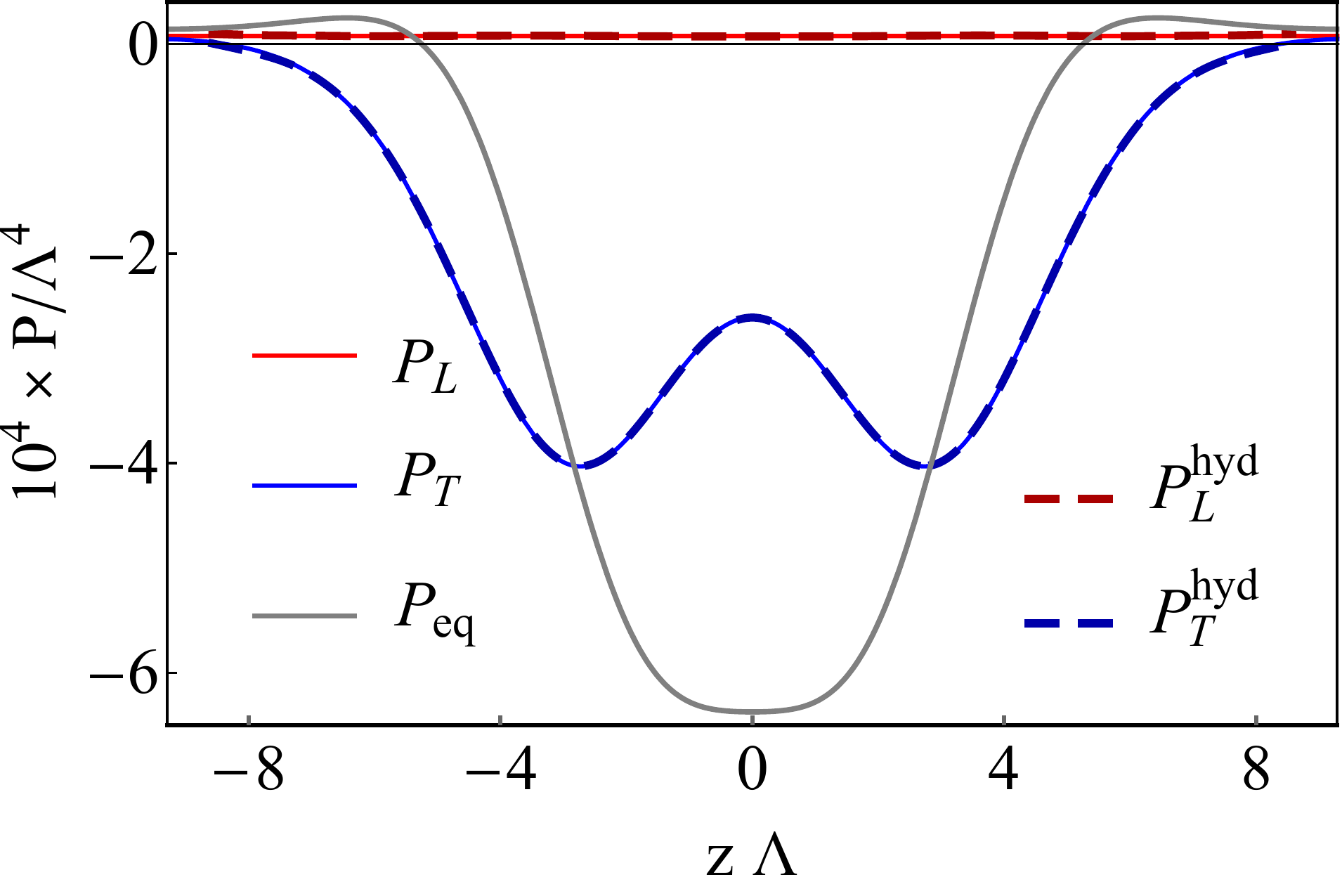} 
\vspace{7mm} \\
\includegraphics[width=.7\textwidth]{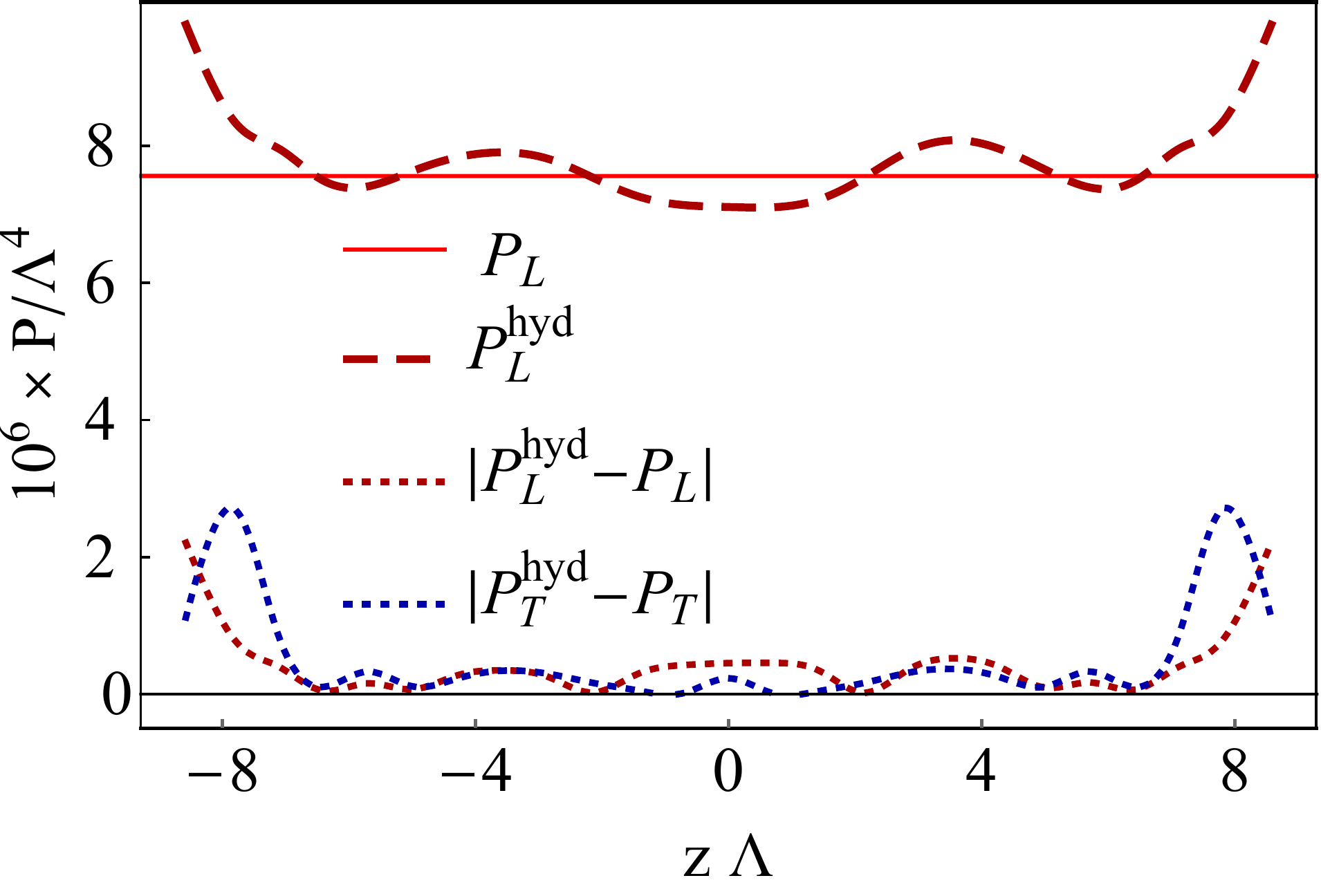} 
\caption{\label{fig:pressures} (Top) Pressures of the end state of \fig{3Denergy}. 
The true pressures deviate significantly from the $\Peq$ but are in excellent agreement with  the hydrodynamic predictions \eqq{hydroap1}-\eqq{hydroap2}. (Bottom) Zoom-in plot of $\Pl$ and difference between the true pressures and the hydrodynamic predictions. Note the different scales in the top and bottom plots. 
}
\end{center}
\end{figure}

Equivalently, a hydrodynamic description of the final states means that, once $c_{\rm L, \,  T} (\E)$ and $f_{\rm L,\, T} (\E)$ have been extracted from a given state (or computed microscopically), they can be used to predict the pressures of a different state given its energy density profile. This is illustrated in \fig{fig:pressures}, where we compare the true pressures obtained from the gravity side for the end state of \fig{3Denergy} with the hydrodynamic predictions  \eqq{hydroap1} and \eqq{hydroap2} based on coefficients extracted from a different state. It is remarkable that hydrodynamics works despite the fact that the second-order terms in these equations are as large as the equilibrium pressure, as can be seen in the figure from the difference between the continuous gray curve and the true pressures. This is particularly dramatic for the longitudinal pressure, $\Pl$, which must be 
$z$-independent since, in a static state, conservation of the stress tensor reduces to $\partial_z T^{zz}=0$. While the equilibrium contribution to $\Pl$ inherits 
$z$ dependence from the energy density, this modulation is precisely compensated by the second-order contribution. We conclude that the second-order gradients sustain the inhomogeneous state. 

Taking the applicability of hydrodynamics one step further, we can use it to predict all other static, inhomogeneous configurations.\footnote{Unless the second-order approximation breaks down for gradients even larger than those that we have considered.} 
Time independence implies that $P_L$ must be constant, which in the hydrodynamic approximation reduces to a second-order, non-linear differential equation for $\E (z)$ via \eqn{hydroap1}. This depends on two integration constants that can be traded for the length of the circle (more precisely, the size of the domains) and the average energy density. Once  $\E (z)$  is known $P_T$ is predicted by \eqn{hydroap2}.

\section{Hydrodynamic evolution}
The second-order coefficients that we have extracted also control the dynamical evolution of the instability. However, unlike in the stationary final configuration, during the dynamical evolution there are non-zero momentum fluxes, leading to a small but non-vanishing three-velocity field. 
 To approximate the full time evolution by hydrodynamics it is thus necessary to include the first-order shear and bulk tensor contributions in \eqn{eq:hydroT}. 
 Similarly, while the systems evolves there are additional second-order gradients beyond those considered in Eqs.~\eqq{hydroap1} and \eqref{hydroap2}. 
 However, these additional  second-order terms are quadratic in the velocity field and can be consistently neglected. 

\begin{figure}[tbhp]
\begin{center}
\includegraphics[width=.7\textwidth]{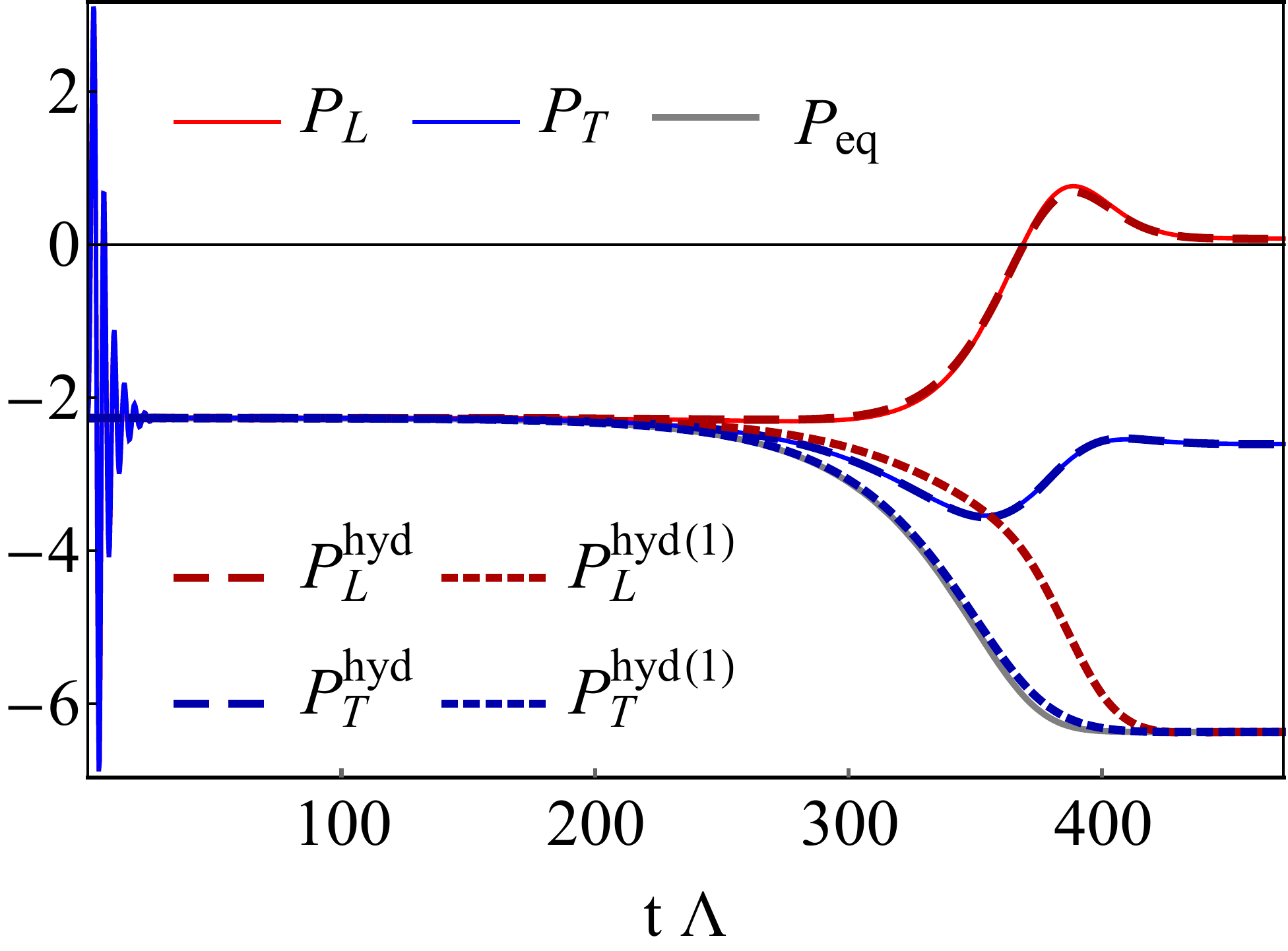}
\caption{\label{fig:poft} 
Time evolution of the pressures at $\Lambda z=0$, in units of $\Lambda^4/10^4$.
$P^{\rm{hyd} (1)}$ denotes the hydrodynamic pressure with only first-order terms included, whereas $P^{\rm{hyd}}$ includes these plus the second-order terms of Eqs.~\eqq{hydroap1} and \eqn{hydroap2}.
}
\end{center}
\end{figure}

 In \fig{fig:poft} we show the time evolution of the pressures at 
 $z \, \Lambda=0$.
 The very early time behavior is the exponential decay of the quasi-normal modes (QNM) that are excited by the perturbation that we introduce to trigger the instability. 
 After this short time, the predictions of  second-order hydrodynamics match the true pressures. 
Second-order terms become increasingly important as the instability saturates, where the first-order approximation, let alone the equilibrium pressure, fails to predict the true pressures while the second-order approximation continues to do so accurately. 
We  conclude that hydrodynamics with large second-order gradients  describes the evolution 
and the saturation of the spinodal instability.

\section{Discussion}
We have uncovered a new example of the applicability of hydrodynamics to  systems with large gradients. As in other known examples, part of the reason behind this success is the relaxation of the non-hydrodynamic QNMs at the relevant time scales of the evolution, as in the early transient behaviour observed in \fig{fig:poft}. The analysis 
of the exponential decay of these modes reveals that this process occurs over a time $\Gamma_{\rm QNM}\sim c\, \pi T$ with $c\simeq 3.4$, consistent with \cite{Buchel:2015saa}. In contrast, for the unstable hydrodynamic mode, assuming $\zeta/\eta \sim 1$ and $\eta/s=1/4\pi$, \eqn{eq:growth}  yields  the typical growth rate $\Gamma\sim \left| c_s\right|^2 \pi T$, which in our case is suppressed with respect to 
$\Gamma_{\rm QNM}$ by the small value of $\left| c_s \right |^2 \simeq 0.03$. Although this argument may explain why hydrodynamics provides a good description at intermediate times, once the QNMs have decayed, it remains surprising that it also describes the late-time evolution and the final state, where the spatial gradients are large. Moreover, it suggests that it would be interesting to explore other values of the parameter $\phi_M$ for which the first-order transition becomes stronger and $\left | c_s\right |$ can become of order unity.

Our static inhomogeneous configurations do not describe the separation of the system into two stable phases. For example, the maxima of the end state of \fig{3Denergy} do not lie on the green stable branch but on the red unstable branch of \fig{fig:energydensity}. Presumably the reason is that, if the available average energy density is small, the cost of the necessary gradients for the peaks to reach the green stable branch makes the corresponding configuration disfavored.  Similar considerations seem to be responsible for the fact that the final state of \fig{3Denergy} exhibits three identical domains. We have found another static configuration with five peaks which is also well described by hydrodynamics but whose entropy is smaller than that of the three-peak state. Moreover, we have studied the dynamics of the system when perturbed by an $n=1$ mode instead of by an $n=3$ mode. Despite the fact that we have followed the evolution for very long times we have  found that  this configuration does not seem to settle down to a stationary state; nevertheless, we have observed that throughout its non-linear evolution it develops a second peak. This suggest that a phase separated configuration that naively one would expect to reach by starting with the $n=1$ mode may simply not exist for the average energy density and the box size that we have considered. A logical possibility compatible with all the above  is therefore that, under these conditions, the three-peak  configuration that we have presented  is the thermodynamically preferred state. We emphasize, however, that establishing this definitively would require comparing the entropies of \emph{all} possible inhomogeneous states of the system with the same average energy density and box size,  which is certainly beyond the scope of this paper. 

We suspect that the reason why the phase-separated configuration is seemingly not allowed for the box size that we have considered is the fact that the two stable phases in our phase diagram in \fig{fig:energydensity} differ by several orders of magnitude. This suggests that, in order to realise the phase-separated configuration, one may need a much larger box.

As pointed out above, from the gravity viewpoint the gauge theory spinodal instability translates into an instability of the dual black brane. This is similar to the Gregory-Laflamme instability of a black string in five-dimensional flat space \cite{Gregory:1993vy} in that it is a long wavelength instability associated to the sound mode of the system that appears below a certain mass or energy density. However, there is also one crucial dissimilarity that makes  the time evolution and the final state of our system and those of a black string \cite{Lehner:2010pn} completely different. This is the fact that the black string is unstable for any mass density below a certain critical value, whereas in our case the black brane is only unstable in a certain energy band (the red dashed curve in \fig{fig:energydensity}). In the case of the black string, at intermediate stages in the evolution the horizon can be described as a sequence of three-dimensional spherical black holes joined by black string segments. Since these segments are themselves subject to a Gregory-Laflamme instability, the evolution results in a self-similar cascade, where ever-smaller satellite black holes form connected by ever-thinner string segments. In contrast, in our case this  evolution stops (roughly speaking) once the energy density at a certain point is low enough to lie in the stable region of the phase diagram, i.e.~below the dashed red curve in \fig{fig:energydensity}.

\begin{acknowledgements}
We thank the MareNostrum supercomputer at the BSC for computational resources (project no.~UB65). Computations were also performed on the Baltasar cluster at IST. We thank  T~. Andrade, O.~Dias,  R.~Emparan, C.~Hoyos, S.~Krippendorf, K.~Rajagopal, J.~Santos, A.~Starinets and W.~van der Schee for discussions. 
MA is supported by the Marie Sklodowska-Curie Individual Fellowship 
658574 FastTh. JCS is a Royal Society University Research Fellow. MZ acknowledges support through the FCT (Portugal) IF programme, IF/00729/2015. We are also supported by grants MEC FPA2013-46570-C2-1-P, MEC FPA2013-46570-C2-2-P, MDM-2014-0369 of ICCUB, 2014-SGR-104, 2014-SGR-1474, CPAN CSD2007-00042 Consolider-Ingenio 2010, and ERC Starting Grant HoloLHC-306605.

\end{acknowledgements}



\end{document}